\documentclass{article}

\usepackage{arxiv}

\usepackage[utf8]{inputenc} 
\usepackage[T1]{fontenc}    
\usepackage{hyperref}       
\usepackage{url}            
\usepackage{booktabs}       
\usepackage{amsmath}
\usepackage{amsfonts}       
\usepackage{amssymb}
\usepackage{nicefrac}       
\usepackage{microtype}      
\usepackage{cleveref}       
\usepackage{lipsum}         
\usepackage{graphicx}
\usepackage{natbib}
\usepackage{doi}
\usepackage{bbm}

\usepackage[table]{xcolor}  
\usepackage{enumitem}
\usepackage{graphicx}
\usepackage{xspace}
\usepackage{array}
\usepackage{changepage,threeparttable} 
\usepackage{multirow}
\usepackage{caption}
\usepackage{subcaption}

\graphicspath{{figures/}}

\hypersetup{
    colorlinks,
    linkcolor={red!50!black},
    citecolor={blue!50!black},
    urlcolor={blue!80!black}
}

\title{GPS++: An Optimised Hybrid MPNN/Transformer for Molecular Property Prediction}

\date{}

\newcommand{\loc}[1]{\fontsize{9.5pt}{9.5pt}{#1}}
\newcommand{\email}[1]{\fontsize{9.5pt}{9.5pt}{\texttt{#1}}}

\author{
	Dominic Masters \\
	\loc{Graphcore, UK} \\
	\email{dominicm@graphcore.ai} \\
    \And
	Josef Dean \\
	\loc{Graphcore, London, UK} \\
	\email{josefd@graphcore.ai} \\
    \And
	Kerstin Klaser \\
	\loc{Graphcore, London, UK} \\
	\email{kerstink@graphcore.ai} \\
    \And
	Zhiyi Li \\
	\loc{Graphcore, Cambridge, UK} \\
	\email{zhiyil@graphcore.ai} \\
    \And
	Sam Maddrell-Mander \\
	\loc{Graphcore, Bristol, UK} \\
	\email{samuelm@graphcore.ai} \\
    \And
	Adam Sanders \\
	\loc{Graphcore, Bristol, UK} \\
	\email{adams@graphcore.ai} \\
    \And
	Hatem Helal \\
	\loc{Graphcore, Cambridge, UK} \\
	\email{hatemh@graphcore.ai} \\
    \And
	Deniz Beker \\
	\loc{Graphcore, Bristol, UK} \\
	\email{denizb@graphcore.ai} \\
    \And
    Ladislav Rampášek \\
    \loc{Mila, Université de Montréal} \\
    \email{ladislav.rampasek@mila.quebec} \\
    \And
    Dominique Beaini \\
    \loc{Valence Discovery, Mila, Université de Montréal} \\
    \email{dominique@valencediscovery.com} \\
}

\renewcommand{\headeright}{Technical Report}
\renewcommand{\undertitle}{First Place Solution to the 2022 Open Graph Benchmark - Large Scale Challenge}
\renewcommand{\shorttitle}{GPS++}

\begin{document}

\newcommand{\gps}{\texttt{GPS++}\xspace}
\newcommand{\mpnn}{\texttt{MPNN}\xspace}
\newcommand{\attn}{\texttt{BiasedAttn}\xspace}
\newcommand{\mlp}{\texttt{MLP}\xspace}
\newcommand{\ffn}{\texttt{FFN}\xspace}
\newcommand{\dropout}{\texttt{Dropout}\xspace}
\newcommand{\graphdropout}{\texttt{GraphDropout}\xspace}
\newcommand{\dense}{\texttt{Dense}\xspace}
\newcommand{\gelu}{\texttt{GELU}\xspace}
\newcommand{\relu}{\texttt{ReLU}\xspace}
\newcommand{\layernorm}{\texttt{LayerNorm}\xspace}
\newcommand{\encoder}{\texttt{Encoder}\xspace}
\newcommand{\decoder}{\texttt{Decoder}\xspace}
\newcommand{\embed}{\texttt{Embed}\xspace}
\newcommand{\softmax}{\texttt{Softmax}\xspace}
\def\xmathbf#1{\textcolor{red}{#1}}
\newcommand{\X}{\xmathbf{x}_i}
\newcommand{\Y}{\xmathbf{y}_i}
\newcommand{\Z}{\xmathbf{z}_i}
\newcommand{\h}{\xmathbf{h}_i}
\newcommand{\x}{\xmathbf{x}}
\newcommand{\y}{\xmathbf{y}}
\newcommand{\z}{\xmathbf{z}}
\newcommand{\e}{\xmathbf{e}}
\newcommand{\R}{\xmathbf{r}}
\newcommand{\E}{\xmathbf{e}_{uv}}
\newcommand{\m}{\xmathbf{m}_{uv}}
\newcommand{\M}{\xmathbf{m}_{uv}}
\newcommand{\A}{\xmathbf{A}}
\newcommand{\B}{\xmathbf{B}}
\newcommand{\curlyE}{\mathcal{E}}
\newcommand{\curlyV}{\mathcal{V}}
\newcommand{\curlyG}{\mathcal{G}}
\newcommand{\curlyX}{\mathcal{X}}
\newcommand{\curlyNu}{\mathcal{N}_u(i)}
\newcommand{\curlyNv}{\mathcal{N}_v(i)}

\definecolor{darkgreen}{rgb}{0, 0.6, 0} 

\newcommand{\lr}[1]{{{\textcolor{purple}{\textbf{[LR:} {#1}\textbf{]}}}}}
\newcommand{\dm}[1]{{{\textcolor{blue}{\textbf{[DM:} {#1}\textbf{]}}}}}
\newcommand{\josefd}[1]{{{\textcolor{red}{\textbf{[JD:} {#1}\textbf{]}}}}}
\newcommand{\unfinished}[1]{{{\textcolor{red}{\textbf{[unfinished:} {#1}\textbf{]}}}}}
\newcommand{\db}[1]{{{\textcolor{darkgreen}{\textbf{[DB:} {#1}\textbf{]}}}}}
\newcommand{\awf}[1]{{{\textcolor{orange}{\textbf{[AWF:} {#1}\textbf{]}}}}}
\newcommand{\hh}[1]{{{\textcolor{purple}{\textbf{[HH:} {#1}\textbf{]}}}}}
\newcommand{\kk}[1]{{{\textcolor{magenta}{\textbf{[KK:} {#1}\textbf{]}}}}}

\maketitle

\begin{abstract}
This technical report presents GPS++, the first-place solution to the Open Graph Benchmark Large-Scale Challenge (OGB-LSC 2022) for the PCQM4Mv2 molecular property prediction task. Our approach implements several key principles from the prior literature. At its core our GPS++ method is a hybrid MPNN/Transformer model that incorporates 3D atom positions and an auxiliary denoising task. The effectiveness of GPS++ is demonstrated by achieving 0.0719 mean absolute error on the independent \texttt{test-challenge} PCQM4Mv2 split. Thanks to Graphcore IPU acceleration, GPS++ scales to deep architectures (16 layers), training at 3 minutes per epoch, and large ensemble (112 models), completing the final predictions in 1 hour 32 minutes, well under the 4 hour inference budget allocated. Our implementation is publicly available at: \url{https://github.com/graphcore/ogb-lsc-pcqm4mv2}.
\end{abstract}

\keywords{Molecular property prediction, Graph learning, Hybrid MPNN/Transformer, OGB-LSC PCQM4Mv2}

\section{Introduction}
In a push to accelerate development of machine learning on graph-structured data, Open Graph Benchmark (OGB)~\citep{hu2020ogb} was created with a variety of graph learning tasks in mind, ranging from graph-level prediction, to link-level prediction, to node-level prediction tasks. Each of these categories has its own challenges, particularly when scaling the application to considerably larger sets of graphs or to graphs with a considerably larger number of nodes. The application at hand and the desired scaling direction thus majorly impacts the machine learning method development. To encourage development of methods for highly impactful applications, OGB Large Scale Challenge (LSC)~\citep{hu2021ogblsc} was organised and for the first time held at KDD 2021. In this technical report we present our \gps submission to OGB-LSC 2022, the second installment of the challenge, for the graph-level prediction task PCQM4Mv2.

The PCQM4Mv2 dataset~\citep{hu2021ogblsc} is specifically aimed at aiding the development of machine learning methods for molecular property prediction. The task presented is to predict the HOMO-LUMO energy gap of a molecule, a property that is typically calculated using Density Functional Theory (DFT) \citep{kohn_sham_1965}. DFT is the de facto method used for accurately predicting quantum phenomena across a range of molecular systems. Unfortunately, traditional DFT can be extremely computationally expensive, prohibiting the efficient exploration of chemical space \citep{dobson_chemical_2004}. Within this context the motivation for replacing it with fast and accurate machine learning models is clear. While this task does aim to accelerate the development of new methods for DFT it also serves as a proxy for other molecular property prediction tasks. It therefore has the potential to benefit a wide range of scientific applications in fields like computational chemistry, material sciences and drug discovery.

In this work we present \gps, a hybrid message passing neural network (MPNN) and transformer that builds on the General, Powerful, Scalable (GPS) framework presented by \cite{rampasek2022GPS}. Specifically, we combine a large and expressive message passing module with a biased self-attention layer to maximise the benefit of local inductive biases while still allowing for effective global communication. Furthermore, we integrate a grouped input masking method to exploit available 3D positional information and use a denoising loss to alleviate oversmoothing.

We accelerate the training and inference of \gps model with Graphcore IPUs allowing us to train our final 44M parameter model in under 24 hours, just 3 minutes per epoch. This final model achieves comparative single-model mean absolute error (MAE) to the state of the art transformers with only 60\% of the parameters.

Our final competition submission consists of a 112 model ensemble, which due to hardware acceleration completes inference in 1 hour and 32 minutes, well under the 4 hour budget allowed. We achieve a final MAE of 0.0719 on the \texttt{test-challenge} data split achieving a top-3 position in the competition. 

\section{Preliminaries} \label{sec:dataset}

\subsection{Notation}
\def\RR#1{\mathbb{R}^{#1}}
\def\brak#1{\bigl[#1\bigr]}
\def\hcat{~\mid~} \def\hcatname{a vertical bar}
\def\Bigghcat{~~~\Bigg|~}
\def\vcat{~;~} \def\vcatname{a semicolon}
\def\vfor{~\mathsf{for}~}
\def\vv{\mathbf v}
Throughout the paper we use the following notation.
Bold lowercase letters $\vv$ are (row) vectors, bold uppercase letters $\mathbf M$ are matrices, with individual elements denoted by non-bold letters i.e.\ $v_k$ or $M_{pq}$.
Blackboard bold lowercase letters~$\mathbbm v$ are categorical (integer-valued) vectors. 
In general, we denote by $\brak{\vv_k}_{k\in K}$ the vertical concatenation (stacking) of vectors~$\vv_k$.
Vertical concatenation is also denoted by \vcatname, i.e. $\brak{\vv_1 
\vcat ... \vcat \vv_J} = \brak{\vv_j}_{j = 1}^J = \brak{\vv_j \vfor j \in \{1..J\}}$.
Horizontal concatenation, which typically means concatenation along the feature dimension, is denoted by \hcatname, i.e. $\brak{\vv_1 \hcat \vv_2}$.

\def\myforall #1:{\forall #1:~~~} 

\subsection{Dataset}

The PCQM4Mv2 dataset~\citep{hu2021ogblsc} consists of 3.7M molecules defined by their SMILES strings.  Each molecule can be represented as a graph $\curlyG = (\curlyV, \curlyE)$ for nodes $\curlyV$ and edges $\curlyE$. In this representation, each node $i\in \curlyV$ is an atom in the molecule and each edge $(u, v)\in \curlyE$ is a chemical bond between two atoms.  The number of atoms in the molecule is denoted by~$N = |\curlyV|$ and the number of edges is $M=|\curlyE|$. Each molecule has on average 14 atoms and 15 chemical bonds. However, as the bonds are undirected in nature and graph neural networks act on directed edges, two bidirectional edges are used to represent each chemical bond. 

The 3.7M molecules are separated into standardised sets by OGB \cite{hu2021ogblsc}, namely into \texttt{training} (90\%), \texttt{validation} (2\%), \texttt{test-dev} (4\%) and \texttt{test-challenge} (4\%) sets using a scaffold split where the HOMO-LUMO gap targets are only publicly available for the \texttt{training} and \texttt{validation} splits.

\def\xin{\mathbbm{x}} 
\def\ein{\mathbbm{e}}
\def\mR{\mathbf{R}}
\def\vr{\mathbf{r}}
Each node and edge is also associated with a list of categorical features
$\xin_i \in \mathbb{Z}^{D_{\text{atom}}}$ and
$\ein_{uv} \in \mathbb{Z}^{D_{\text{bond}}}$, respectively, for $D_{\text{atom}}$ atom features and $D_{\text{bond}}$ bond features. A further set of 3D atom positions $\mR = \brak{\vr_1 \vcat ... \vcat \vr_N}\in \mathbb{R}^{N\times 3}$, extracted from the original DFT simulations, is also provided for the training data, but crucially not for the validation and test data.

\def\l{^\ell}

\def\vx{\mathbf{x}}
\def\mX{\mathbf{X}}
\def\ve{\mathbf{e}}
\def\mE{\mathbf{E}}
\def\vg{\mathbf{g}}
\def\xdim{{d_\text{node}}}
\def\edim{{d_\text{edge}}}
\def\gdim{{d_\text{global}}}
Our algorithm operates on edge, node, and global {\em features}.
Node features in layer $\ell$ are denoted by $\vx_i^\ell \in \RR\xdim$,
and are concatenated into the $N \times \xdim$ matrix 
$\mX^\ell = \brak{\vx^\ell_1 \vcat ... \vcat \vx_N^\ell}$.
Edge features $\ve_{uv}^\ell \in \RR\edim$ are concatenated into the edge feature matrix
$\mE^\ell = \brak{\ve^\ell_{uv} \vfor (u,v) \in \curlyE}$.
Global features are defined per layer as $\vg^\ell \in \RR\gdim$.

In this work, we set $\xdim=256$, $\edim=128$ and $d_\textrm{global}=64$.

\def\mB{\mathbf{B}}
We also define an {\em attention bias} matrix $\mB \in \RR{N\times N}$, computed from the input graph topology and 3D atom positions, described later in Section \ref{sec:input_features}.

\section{GPS++}
Our \gps model closely follows the GPS framework set out in \cite{rampasek2022GPS}. 
This work presents a flexible model structure for building hybrid MPNN/Transformer models for graph-structured input data. 
We build a specific implementation of GPS that focuses on maximising the benefit of the inductive biases of the graph structure and 3D positional information. We do this by building a large and expressive MPNN component and biasing our attention component with structural and positional information. We also allow global information to be propagated through two mechanisms, the global attention and by using a global feature in the MPNN.

The main \gps block (Section~\ref{sec:gps_block}) combines the benefits of both message passing and attention layers by running them in parallel before combining them with a simple summation and MLP; this is repeated $16$ times.
This main trunk of processing is also preceded by a an \encoder function (Section \ref{sec:input_features}) responsible for encoding the input information into the latent space and followed by a simple \decoder function (Section \ref{sec:training}).

Feature engineering is also used to improve the representation of the atoms/bonds, to provide the  rich positional and structural features that increase expressivity, and to bias the attention weights with a distance embedding.

\subsection{GPS++ Block} \label{sec:gps_block}

\def\vy{\mathbf{y}}
\def\mY{\mathbf{Y}}
\def\vz{\mathbf{z}}
\def\mZ{\mathbf{Z}}

The $\gps$ block is defined as follows for layers $\ell > 0$ (see Section~\ref{sec:input_features} for the definitions of $\mX^0, \mE^0, \vg^0$).
\begin{align}\label{eqn:gps_layer}
    \mX^{\ell+1}, \mE^{\ell+1}, \vg^{\ell+1} &= \gps \left( \mX\l, \mE\l, \vg\l, \mB \right) \\
    \textrm{computed as} \quad
    \mY\l, \ \mE^{\ell+1}, \ \vg^{\ell+1} &= \mpnn \left(\mX\l, \mE\l, \vg\l \right),\\
    \mZ\l &=  \attn \left(\mX\l, \mB  \right),\\
    \myforall i: \vx_i^{\ell+1} &= \ffn\left(\vy_i\l + \vz_i\l\right)
\end{align}

\begin{figure}[t]
  \centering
    \includegraphics[width=1.0\textwidth]{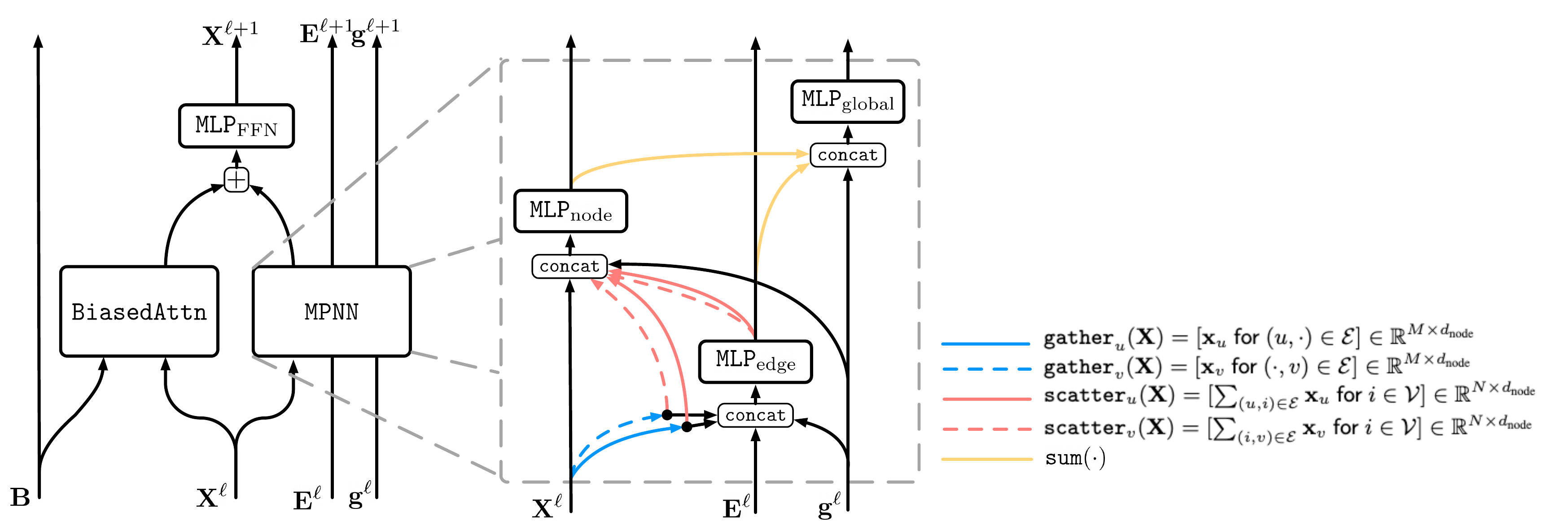}
  \caption{The main \gps processing block (left) is composed of a local message passing \mpnn module (right) and a biased global attention \attn module. The right part of the figure shows a diagram of the used \mpnn block. The \texttt{gather}, \texttt{scatter} and \texttt{sum} operations highlight changes in tensor shapes and are defined as above.}
\label{fig:GPSplusplus}
\end{figure}

\paragraph{The \mpnn module} Our \mpnn module is a variation on the neural message passing module with edge and global features \citep{gilmer2017neural,battaglia2018graphnets,bronstein2021geometric}. We choose this form to maximise the expressivity of the model with the expectation that over-fitting will be less of an issue with PCQM4Mv2, compared to other molecular datasets, due to its size. This \mpnn module is defined as follows (see Figure \ref{fig:GPSplusplus} for a graphical representation):
\begin{align}\label{eqn:mpnn_layer}
    \mY\l, \ \mE^{\ell+1}, \ \vg^{\ell+1} &= \mpnn \left(\mX\l, \mE\l, \vg\l \right), \\
    \textrm{computed as} \qquad
    \myforall (u,v): \bar{\ve}_{uv}\l &=
    \dropout_{0.0035}\left(\mlp_{\textrm{edge}}\left(\brak{
      \vx_u\l \hcat \vx_v\l  \hcat \ve_{uv}\l \hcat \vg\l
    }\right)\right) \label{eqn:message}\\
    \myforall i: \bar{\vx}_{i}\l &= \mlp_{\textrm{node}}\left(\Biggl[
                                                            \vx_i\l
                                                            \Bigghcat 
                                                            \sum_{(u,i)\in\curlyE} \brak{\bar{\ve}_{ui}\l
                                                                                               \hcat
                                                                                               \vx_u\l} 
                                                            \Bigghcat
                                                            \sum_{(i,v)\in\curlyE} \brak{\bar{\ve}_{iv}\l 
                                                                                         \hcat
                                                                                         \vx_v\l}
                                                            \Bigghcat
                                                            \vg\l 
                                                        \Biggr]\right) \label{eqn:node_aggregate}\\
    \bar{\vg}\l &= \mlp_{\textrm{global}}\left(\Biggl[{
                                                    \vg\l
                                                    \Bigghcat
                                                    \sum_{j\in\curlyV} \bar{\vx}_{j}\l
                                                    \Bigghcat
                                                    \sum_{(u,v)\in\curlyE} \bar{\ve}_{uv}\l
                                         }\Biggr]\right) \\
    \myforall i: \vy_i\l &= \layernorm(\dropout_{0.3}(\bar{\vx}_i\l)) + \vx_i\l \\
    \myforall (u,v): \ve_{uv}^{\ell+1} &= \bar{\ve}_{uv}\l + \ve_{uv}\l \\\
    \vg^{\ell+1} &= \dropout_{0.35}(\bar{\vg}\l) + \vg\l,
\end{align}
where $\dropout_p$ \citep{srivastava2014dropout} masks by zero each element with probability $p$ and \layernorm follows the normalisation procedure by \cite{ba2016layernorm}.
The three networks $\mlp_\eta$ for $\eta \in \{\textrm{node}, \textrm{edge}, \textrm{global}\}$ each have two layers and are defined by:
\begin{align}
    \vy &= \mlp_\eta(\vx)\\
    \textrm{computed as} \ \ \ \ 
    \bar{\vx} &= \gelu(\dense(\vx)) \in \mathbb{R}^{4d_\eta} \\
    \vy &= \dense(\layernorm(\bar{\vx})) \in \mathbb{R}^{d_\eta}
\end{align}
where \gelu is an activation function defined in \cite{hendrycks2016gelu}.

This message passing block is principally the most similar to \cite{battaglia2018graphnets}. Our variation is predominantly two fold: i) we concatenate the node representations to the incident edge in the formation of ``messages'', and ii) we concatenate inputs to the $\mlp$ rather than sum.

\paragraph{The \attn module} Our \attn module follows the form of a biased self attention by \cite{ying2021graphormer} where a standard self attention block \citep{vaswani2017attention} is biased by a structural prior derived from the input graph. In our work the bias $\mB$ is made up of two components, a shortest path distance embedding and a 3D distance bias derived from the molecular conformations as described in Section~\ref{sec:input_features}. Single-head attention is defined as:
\def\mW{\mathbf W}
\begin{align}
\attn(\mathbf{X}, \mB) &=  \textrm{\softmax}\left( \frac{\left( \mathbf{X} \mW_Q \right) \left( \mathbf{X} \mW_K \right)^\top}{\sqrt{\xdim}} + \mB \right) \left( \mathbf{X} \mW_V \right) \in\RR{N\times \xdim}
\end{align}
for learnable weight matrices $\mW_Q, \mW_K, \mW_V \in \mathbb{R}^{\xdim\times\xdim}$, though in practice we use 32 heads. 

\paragraph{The \ffn module}
Finally, the feed-forward network module takes the form:
\begin{align}
    \vy &= \ffn(\vx)\\
    \textrm{computed as} \ \ \ \ 
    \bar{\vx} &= \dropout_{p}(\gelu(\dense(\vx))  \quad &&\in \mathbb{R}^{4\xdim} \\
    \vy &= \graphdropout_{\frac{\ell}{L}0.3}(\dense(\bar{\vx})) + \vx   \quad &&\in \mathbb{R}^{\xdim}
\end{align}
Unless otherwise stated, the dropout probability $p=0$, however, we experiment with other values when ensembling multiple model variants.

\begin{table}[t]
    \caption{Allocation of the number of parameters (in millions) throughout the model.
        }
    \label{tab:parameter_counts}
    \centering
    \begin{tabular}{lrr}\toprule
    \multicolumn{1}{l}{\textbf{Location}} & \multicolumn{2}{l}{\textbf{\# Parameters (M)}} \\
    \midrule
    Encoder         & 0.63 & 1.4\%                          \\
    $\gps$ (16 blocks)           & \hspace{1em} 43.7 & 98.6\%                           \\
    \quad$\mpnn$            & 31.0 & 70.0\%                           \\
    \quad\quad $\mlp_{\text{node}}$   & 22.1 & 50.0\%                           \\
    \quad\quad $\mlp_{\text{edge}}$   & 6.84 & 15.4\%                           \\
    \quad\quad $\mlp_{\text{global}}$ & 2.11 & 4.8\%                          \\
    \quad\attn      & 4.22 & 9.5\%                          \\
    \quad$\mlp_{\text{FFN}}$             & 8.42 & 19.0\%                           \\
    Decoder         & 0.12 & 0.3\%                          \\
    \midrule
    \textbf{Total}           & \textbf{44.3} & \textbf{100\%}   \\
    \bottomrule                      
    \end{tabular}
    \vspace{-5pt}
\end{table}

Table~\ref{tab:parameter_counts} shows how the learnable weights are distributed through the model. Interestingly, 70\% of the total parameters are located in the MPNN, highlighting the focus of this model using local structures and exploiting the inductive bias of the graph.

\subsection{Input Feature Engineering}\label{sec:input_features}

\def\threeD{^\text{3D}}
As described in Section~\ref{sec:dataset}, the dataset samples include the graph structure $\curlyG$, a set of categorical features for the atoms and bonds $\xin_i$, $\ein_{uv}$, and the 3D node positions $\vr_i$. It has been shown that there are many benefits to augmenting the input data with additional structural, positional, and chemical information  \citep{rampasek2022GPS, wang2022equivstable, dwivedi2022LPE}. Therefore, we combine several feature sources when computing the input to the first \gps layer. There are four feature tensors to initialise; node state, edge state, whole graph state and attention biases.
\begin{align}
    \mX^{0}&=\dense(\brak{\mX^\text{atom} \hcat \mX^\text{LapVec} \hcat \mX^\text{LapVal} \hcat \mX^\text{RW} \hcat \mX^\text{Cent} \hcat \mX\threeD}) &&\in\RR{N\times \xdim}\\
    \mE^{0}&= \dense(\brak{\mE^\text{bond} \hcat \mE\threeD}) &&\in\RR{M\times \edim}\\
    \vg^0 &= \embed_{\gdim}(0) &&\in\RR{\gdim}\\
    \mB&=\mB^{\mathrm{SPD}} + \mB\threeD &&\in\RR{N\times N}
\end{align}
The various components of each of these equations are defined over the remainder of this section.
The encoding of these features also makes recurring use of the following two generic functions. Firstly, a two-layer $\mlp_{\mathrm{encoder}}$ that projects features to a fixed-size latent space:
\begin{align}
    y &= \mlp_{\textrm{encoder}}(x), \quad\text{where}\, x\in\RR{h}\\
    \textrm{computed as} \ \ \ \ 
    \bar{x} &= \relu(\dense(\layernorm(x))) &&\in \mathbb{R}^{2h} \\
    y &= \dropout_{0.18}(\dense(\layernorm(\bar{x}))) &&\in \mathbb{R}^{32}
\end{align}
Secondly, a function $\embed_{d}(j)\in\RR{d}$ which selects the $j^\mathrm{th}$ row from an implicit learnable weight matrix.

\paragraph{Chemical Features} The categorical features $\xin$, $\ein$ supplied with the original dataset represent a set of 9~atom and  3~bond features (described in Table~\ref{tab:chemical_input_features}). There are, however, a wide range of chemical features that can be extracted from the periodic table or using tools like \cite{rdkit}. \cite{ying2021first} have shown that extracting additional atom level properties can be beneficial when trying to predict the HOMO-LUMO energy gap, defining a total of 28 atom and 5 bond features. We explore the impact of a number of additional node and edge features and sweep a wide range of possible combinations. 

In particular, we expand on the set defined by \cite{ying2021first} with three additional atom features derived from the periodic table, the atom group (column), period (row) and element type (often shown by colour). We found that these three additional features were particularly beneficial. Furthermore, we hoped that this would allow us to drop the atomic number, which can be determined from the combination of group and period for all elements occurring in the dataset, enabling the model to generalise to atoms not seen in the training set. However, we found that atomic number was important to keep even in the presence of these additional features.

We also found that in many cases \emph{removing} features was beneficial, for example, we found that all our models performed better when excluding information about chiral tag and replacing it by chiral centers. We further observe that our best feature combinations all consist of only 8 node features, where the majority of the input features stay consistent between the sets. 
We show the three best feature sets found in Table~\ref{tab:chemical_input_features} and use \emph{Set 1} for all experiments unless otherwise stated (e.g., during ensembling).

Finally, to embed the categorical chemical features from the dataset $\xin_i$, $\ein_{uv}$ into a continuous vector space, we learn a simple embedding vector for each category, sum the embeddings for all categories, and then process it with an \mlp to produce $\mX^\text{atom}$ and $\mE^\text{bond}$, i.e.
\begin{align}
    \myforall i: \vx_i^\text{atom} &=\dropout_{0.18}\left(\mlp_{\textrm{node}}\left(\sum\nolimits_{j \in \xin_i} \embed_{64}(j)\right)\right)\in\RR{\xdim}\\
    \myforall (u, v): \ve_{uv}^\text{bond} &=\dropout_{0.18}\left(\mlp_{\textrm{edge}}\left(\sum\nolimits_{j \in \ein_{uv}} \embed_{64}(j)\right)\right)\in\RR{\edim}
\end{align}
Here $\mlp_\textrm{node}$ and $\mlp_\textrm{edge}$ refer to the functions by the same names used in Eq.~\ref{eqn:message}~and~\ref{eqn:node_aggregate} in the \mpnn module, yet parameterised independently.

\begin{table}
    \caption{Chemical input feature selection for PCQM4Mv2 dataset.
    }
    \label{tab:chemical_input_features}
    \centering
    \fontsize{9pt}{9pt}\selectfont
    \rowcolors{1}{}{lightgray!20}
    \renewcommand{\arraystretch}{1.1}
    \begin{tabular}{lcccc}\toprule
    \textbf{} & \multicolumn{4}{c}{\textbf{Feature Set}}\\ \cmidrule(l{2pt}r{2pt}){2-5}
    \textbf{Node features}&\textbf{Original} &$\textbf{Set 1}$ &\textbf{Set 2} &\textbf{Set 3} \\\midrule
    Atomic number & $\checkmark$ & $\checkmark$ & $\checkmark$ & $\checkmark$ \\
    Group & $\times$ & $\checkmark$ & $\checkmark$ & $\checkmark$ \\
    Period & $\times$ & $\checkmark$ & $\checkmark$ & $\checkmark$ \\
    Element type & $\times$ & $\checkmark$ & $\checkmark$ & $\checkmark$ \\
    Chiral tag & $\checkmark$ & $\times$ & $\times$  & $\times$ \\
    Degree & $\checkmark$ & $\checkmark$ & $\checkmark$  & $\checkmark$ \\
    Formal charge & $\checkmark$ & $\checkmark$ & $\times$  & $\checkmark$ \\
    \# Hydrogens & $\checkmark$ & $\checkmark$ & $\checkmark$  & $\checkmark$ \\
    \# Radical electrons & $\checkmark$ & $\checkmark$ & $\checkmark$  & $\checkmark$ \\
    Hybridisation & $\checkmark$ & $\times$ & $\checkmark$  & $\checkmark$ \\
    Is aromatic & $\checkmark$ & $\checkmark$ & $\checkmark$  & $\times$ \\
    Is in ring & $\checkmark$ & $\checkmark$ & $\checkmark$  & $\checkmark$ \\
    Is chiral center & $\times$ & $\checkmark$ & $\checkmark$  & $\checkmark$ \\
    \textbf{} & \textbf{} & \textbf{} & \textbf{}  & \textbf{} \rowcolors{0}{white}{}\\
    \textbf{Edge features}&\textbf{} &\textbf{} &\textbf{} &\textbf{} \\\midrule
    Bond type & $\checkmark$ & $\checkmark$ & $\times$  & $\times$ \\
    Bond stereo & $\checkmark$ & $\checkmark$ & $\checkmark$  & $\checkmark$ \\
    Is conjugated & $\checkmark$ & $\times$ & $\checkmark$  & $\checkmark$ \\
    Is in ring & $\times$ & $\checkmark$ & $\checkmark$  & $\checkmark$ \\
    
    \bottomrule
    \end{tabular}
    \vspace{-5pt}
\end{table}

\def\mA{\mathbf{A}}
\def\mD{\mathbf{D}}
\def\mL{\mathbf{L}}
\def\mU{\mathbf{U}}
\def\mP{\mathbf{P}}
\def\mLambda{\mathbf{\Lambda}}
\def\diag{\mathrm{diag}}
\paragraph{Graph Laplacian Positional Encodings {\normalfont\citep{kreuzer2021rethinking, dwivedi2020generalization}}} Given a graph with adjacency matrix $\mA$ and degree matrix $\mD$, the eigendecomposition of the graph laplacian $\mL$ is formulated into a global positional encoding as follows.
\begin{align}
    \myforall i: \vx_{i}^{\mathrm{LapVec}} &= \mlp_{\textrm{encoder}} (\mU \left[ i, \, 2 \dots {k^\mathrm{Lap}} \right])\in\RR{32}, \quad\mathrm{ where }\  \mL = \mD - \mA = \mU^\top \mLambda \mU\label{eq:LapVec}\\
    \myforall i: \vx_{i}^{\mathrm{LapVal}} &= \mlp_{\textrm{encoder}} \left(\frac{\mLambda'}{||\mLambda'||}\right)\in\RR{32}, \quad\mathrm{ where }\ \mLambda'=\diag(\mLambda)\left[2\dots k^\mathrm{Lap}\right] \label{eq:LapVal}
\end{align}
To produce fixed shape inputs despite variable numbers of eigenvalues / eigenvectors per graph, we truncate / pad to the lowest $7$ eigenvalues, excluding the first trivial eigenvalue $\Lambda_{11}=0$. We also randomise the eigenvector sign every epoch which is otherwise arbitrarily defined.

\paragraph{Random Walk Structural Encoding {\normalfont\citep{dwivedi2022LPE}}} This feature captures the probability that a random graph walk starting at node $i$ will finish back at node $i$, and is computed using powers of the transition matrix $\mP$. This feature captures information about the local structures in the neighbourhood around each node, with the degree of locality controlled by the number of steps. For this submission random walks from $1$ up to $k^\mathrm{RW}=16$ steps were computed to form the feature vector.
\begin{align}
    \myforall i: \vx_{i}^{\mathrm{RW}} &=\mlp_{\textrm{encoder}} \left(\left[ (\mP^1)_{ii},\, (\mP^2)_{ii},\, \cdots,\, (\mP^{k^\mathrm{RW}})_{ii} \right]\right) \in\RR{32}, \quad\mathrm{ where }\   \mP = \mD^{-1} \mA \label{eq:RandWalk}
\end{align}

\paragraph{Local Graph Centrality Encoding {\normalfont\citep{ying2021graphormer,shi2022benchgraphormer}}} The graph centrality encoding is intended to allow the network to gauge the importance of a node based on its connectivity, by embedding the degree (number of incident edges) of each node into a learnable feature vector.
\begin{align}
    \myforall i: \vx_{i}^\text{Cent} &=\embed_{64} \left(D_{ii}\right)\in\RR{64} \label{eq:NodeCent}
\end{align}

\paragraph{Shortest Path Distance Attention Bias}
Graphormer \citep{ying2021graphormer,shi2022benchgraphormer} showed that graph topology information can be incorporated into a node transformer by adding learnable biases to the self-attention matrix depending on the distance between node pairs. During data preprocessing the Shortest Path Distance (SPD) map $\Delta \in \mathbb{N}^{N\times N}$ is computed where $\Delta_{ij}$ is the number of edges in the shortest continuous path from node $i$ to node $j$. During training each integer distance is embedded as a scalar attention bias term to create the SPD attention bias map $\mB^\mathrm{SPD}\in\mathbb{R}^{N\times N}$.
\begin{align}
    \myforall i,j: B_{ij}^\mathrm{SPD}=\embed_1\left(\Delta_{ij}\right)\in \mathbb{R}
\end{align}
Single-headed attention is assumed throughout this report for simplified notation, however, upon extension to multi-headed attention, one bias is learned per distance per head.

\paragraph{Embedding 3D Distances}
Using the 3D positional information provided by the dataset comes with a number of inherent difficulties. Firstly, the task is invariant to molecular rotations and translations, however, the 3D positions themselves are not. Secondly, the 3D conformer positions are only provided for the training data, not the validation or test data. To deal with these two issues and take advantage of the 3D positions provided we follow the approach of \cite{luo2022transformerM}. 

\def\vpsi{\boldsymbol\psi}
To ensure rotational and translational invariance we use only the distances between atoms, not the positions directly. To embed the scalar distances into vector space $\mathbb{R}^{K}$ we first apply $K=128$ Gaussian kernel functions, where the $k^\text{th}$ function is defined as 
\begin{align}
    \myforall i,j: \psi_{ij}^k = - \frac{1}{\sqrt{2\pi}\left|\sigma^k\right|}
    \exp\left(-\frac{1}{2}\left( \frac{||\vr_i-\vr_j||-\mu^k}{|\sigma^k|} \right)^2 \right)\in\mathbb{R}
\end{align}
with learnable parameters $\mu^k$ and $\sigma^k$.
The $K$ elements are concatenated into vector $\vpsi_{ij}$.
We then process these distance embeddings in three ways to produce attention biases, node features and edge features. 

\paragraph{3D Distance Attention Bias} The 3D attention bias map $\mB\threeD\in\mathbb{R}^{N\times N}$ allows the model to modulate the information flowing between two node representations during self-attention based on the spatial distance between them, and are calculated as per \cite{luo2022transformerM}
\begin{align}
    \myforall i,j: \mB_{ij}\threeD&=\mlp_{\textrm{bias3D}}(\vpsi_{ij}) \in \mathbb{R}
    \label{eqn_bias3D}
\end{align}
Upon extension to multi-headed attention with 32 heads $\mlp_\mathrm{bias3D}$ instead projects to $\RR{32}$.

\paragraph{Bond Length Encoding} Whilst $\mB\threeD$ makes inter-node distance information available to the self-attention module in a dense all-to-all manner as a matrix of simple scalar biases, we also make this information available to the \mpnn module in a sparse but high-dimensional manner as edge features $\mE\threeD = \brak{\ve\threeD_{uv} \vfor (u,v) \in \curlyE}$ calculated as
\begin{align}
    \myforall (u, v): \ve_{uv}\threeD&=\mlp_{\textrm{encoder}}(\vpsi_{uv})\in\mathbb{R}^{32}
\end{align}
\paragraph{Global 3D Centrality Encoding} The 3D node centrality features $\mX\threeD=\left[\vx_1\threeD;\,\dots;\,\vx_N\threeD\right]$ are computed by summing the embedded 3D distances from node $i$ to all other nodes. Since the sum commutes this feature cannot be used to determine the distance to a specific node, so serves as a centrality encoding rather than a positional encoding.
\begin{align}
    \myforall i: \vx_i\threeD&=W\threeD\sum_{j\in \curlyV}\vpsi_{ij}\in\mathbb{R}^{32}
\end{align}
Here $W\threeD \in \mathbb{R}^{K\times 32}$ is a linear projection to the same latent size as the other encoded features.

\section{Experimental Setup}\label{sec:exp_setup}

\subsection{Hardware and Acceleration}
\paragraph{Hardware} We train our models using a Graphcore BOW-POD16 which contains 16 IPU processors, delivering a total of 5.6 petaFLOPS of float16 compute and 14.4 GB of in-processor SRAM which is accessible at an aggregate bandwidth of over a petabyte per second. This compute and memory is then distributed evenly over 1472 tiles per chip. This architecture has two key attributes that enable high performance on GNN and other AI workloads \citep{Bilbrey2022ipugnn}: memory is kept as close to the compute as possible (i.e., using on-chip SRAM rather than off-chip DRAM) which maximises bandwidth for a nominal power budget; and compute is split up into many small independent arithmetic units meaning that any available parallelism can be extremely well utilised. In particular this enables very high performance for sparse communication ops, like gather and scatter, and achieves high FLOP utilisation even with complex configurations of smaller matrix multiplications. Both of these cases are particularly prevalent in MPNN structures like those found in GPS++.

To exploit the architectural benefits of the IPU and maximise utilisation, understanding the program structure ahead of time is key. This means all programs must be compiled end-to-end, opening up a range of opportunities for optimisation but also adding the constraint that tensor shapes must be known and fixed at compile time.

\paragraph{Batching and Packing} To enable fixed tensor sizes with variable sized graphs it is common to \emph{pad} the graphs to the max node and edge size in the dataset. This, however, can lead to lots of compute being wasted on padding operations, particularly in cases where there are large variations in the graph sizes. To combat this it is common to \emph{pack} a number of graphs into a fixed size shape to minimise the amount of padding required, this is an abstraction that is common in graph software frameworks like PyTorch Geometric \citep{FeyLenssen2019PyG} and has been shown to achieve as much as 2x throughput improvement for variable length sequence models \citep{krell2021packing}. Packing graphs into one single large pack, however, has a couple of significant downsides: the memory and compute complexity of all-to-all attention layers is $\mathcal{O}(n^2)$ in the pack size not the individual graph sizes, and allowing arbitrary communication between all nodes in the pack forces the compiler to choose sub-optimal parallelisation schemes for the gather/scatter operations.

To strike a balance between these two extremes we employ a two tiered hierarchical batching scheme that packs graphs into a fixed size but then batches multiple packs to form the micro-batch. We define the maximum pack size to be 60 nodes, 120 edges and 8 graphs then use a simple streaming packing method where graphs are added to the pack until either the total nodes, edges or graphs exceeds the maximum size. This achieves 87\% packing efficiency of the nodes and edges with on average 3.6 graphs per pack, though we believe that this could be increased by employing a more complex packing strategy \cite{krell2021packing}. We then form micro-batches of 8 packs which are pipelined \citep{Huang2018gpipe} over 4 IPUs accumulating over 8 micro-batches and replicated 4 times to form a global batch size of 921 graphs distributed over 16 IPUs.

\paragraph{Numerical Precision} To maximise compute throughput and maximise memory efficiency it is now common practice to use lower precision numerical formats in deep learning \citep{Micikevicius2017fp16}. On Graphcore IPUs using float16 increases the peak FLOP rate by 4x compared to float32 but also makes more effective usage of the high bandwidth on-chip SRAM. For this reason we use float16 for nearly all\footnote{A few operations like the sum of squares the variance calculations are up-cast to float32 by the compiler.} compute but also use float16 for the majority\footnote{A small number of weights are kept in float32 for simplicity of the code rather than numerical stability.} of the weights, this is made possible, without loss of accuracy, by enabling the hardware-level stochastic rounding of values. While we do use a loss scaling \cite{Micikevicius2017fp16} value of 1024 we find that our results are robust to a wide range of choices. We also use float16 for the first-order moment in Adam but keep the second-order moment in float32 due to the large dynamic range requirements of the sum-of-squares.

\subsection{Model Training} \label{sec:training}
\paragraph{Training Configuration} Our model training setup uses the Adam optimiser \citep{kingma2015adam} with a gradient clipping value of 5, a peak learning rate of 4e-4 training for a total of 450 epochs. We used a learning rate warmup period of 10 epochs followed by a linear decay schedule.

\paragraph{Decoder and Loss} The final model prediction is formed by global sum-pooling of all node representations and then passing it through a 2-layer MLP. The regression loss is the mean absolute error (L1 loss) between a scalar prediction and the ground truth HOMO-LUMO gap value.

\paragraph{Noisy Nodes/Edges} Noisy nodes \citep{godwin2022noisynodes,zaidi2022pretraining} has previously been shown to be beneficial for molecular GNNs including on the PCQM4M dataset. The method adds noise to the input data then tries to reconstruct the uncorrupted data in an auxilliary task. Its benefits are claimed to be two-fold: it adds regularisation by inducing some noise on the input, but also combats over-smoothing by forcing the node level information to remain discriminative throughout the model. This has been shown to be particularly beneficial when training deep GNNs \citep{godwin2022noisynodes}. We follow the method of \cite{godwin2022noisynodes} that applies noise to the categorical node features by randomly choosing a different category with probability $p_\text{corrupt}$ but also extend this to the categorical edge features, too. A simple categorical cross entropy loss is then used to reconstruct the uncorrupted features at the output. We set $p_\text{corrupt}=0.01$ and weight the cross-entropy losses such that they have a ratio {1:1.2:1.2} for losses {\texttt{HOMO-LUMO}:\texttt{NoisyNodes}:\texttt{NoisyEdges}}.

\paragraph{Grouped Input Masking}
As described in Section~\ref{sec:dataset} the 3D positional features $\mR$ are only defined for the training data. We must therefore make use of them in training without requiring them for validation/test.  We found that the method proposed by \cite{luo2022transformerM} achieved the most favourable results so we adopt a variation hereon referred to as \emph{grouped input masking}.

This method stochastically masks out any features derived from the 3D positional features $\mR$ to build robustness to their absence. Specifically, this is done by defining two input sets to be masked:
\begin{align}
    \curlyX^\text{Spatial}=\lbrace \mX\threeD, \mE\threeD, \mB\threeD \rbrace, \quad \curlyX^\text{Topological}=\lbrace \mB^\mathrm{SPD} \rbrace,
\end{align}
and three potential masking groups: 1. Mask $\curlyX^\text{Spatial}$, 2. Mask $\curlyX^\text{Topological}$, and 3. No masking.
These masking groups are then sampled randomly throughout training with ratio 1:3:1. If 3D positions are not defined, for example in validation/test, masking group 1 is always used.

\paragraph{Training Time} Our \gps model trains at 17500 graphs per second on 16 IPUs which means each epoch completes in 3 minutes and a full 450 epoch training run takes 24 hours.

\paragraph{Dataset Splits} While the original training set already contains 98\% of the labelled data and all of the data with 3D positional information, we still aim to train on all available data for our final model. This, however, comes with many pitfalls due to inability to calculate a reliable measure of performance. Therefore, to understand the value of this additional data we consider an intermediate split configuration where we randomly sample half of the validation set (not resampled per run) to train on, holding the other half out for validation. As a result we have three dataset splits to consider: \texttt{original}, \texttt{train+valid} and \texttt{train+half\_valid}.

\paragraph{Ensembling} Ensembling models has long been used to improve generalisation of machine learning models and has become an indispensable tool for practitioners entering machine learning competitions. Here we outline our ensembling strategy that aims to achieve confidence while training without a validation set, but also allows on-the-fly tuning and weighting of model ensembles. The main idea is to build two comparative model sets to ensemble, a \emph{proxy} set and a \emph{main} set. The proxy set is designed to be qualitatively similar to the main set but maintain a clean held out validation set by training on \texttt{train+half\_valid}, whereas the main set is trained on \texttt{train+valid} . We aim for the proxy set to match the main set in all aspects apart from the training data and the number of models; to be able to focus computational resources on the main set we aim to build the proxy set to be approximately 25\% of its size.

\section{Results}
\paragraph{Single Model Performance} 
In Table~\ref{tab:results_pcqm4m} we compare the single model performance of our model with prior work. It shows that we achieve comparable performance to the best Transformer only model \citep{luo2022transformerM} with just 64\% of the parameters. We believe that this efficiency is driven by the strong inductive bias in the message passing layers which constitute a large proportion of the compute, as also noted in \citet{rampasek2022GPS} when using an MPNN and in \citet{kreuzer2021rethinking} when biasing the attention towards direct neighbours. Furthermore, throughout testing we found that without the inclusion of the 3D distance matrix the self-attention layers often only provided minimal improvement. We therefore hypothesise that one of the main strengths of the attention in this scenario is that it is a highly effective place to integrate the three dimensional distance matrix. We plan to further investigate these points in later revisions of this work.

\begin{table}
    \vspace{-10pt}
    \caption{Comparing single model performance on PCQM4Mv2 dataset.
    }
    \label{tab:results_pcqm4m}
    \centering
    \fontsize{9pt}{9pt}\selectfont
    \begin{tabular}{lccc}\toprule
    \textbf{Model}&\textbf{Model Type}&\textbf{Validation MAE $\downarrow$} &\textbf{\# Param.} \\\midrule
    GRPE~\citep{park2022GRPE} & Transformer &0.0890  &46.2M \\
    EGT~\citep{hussain2022EGT} & Transformer &0.0869  &89.3M \\
    Graphormer~\citep{shi2022benchgraphormer} & Transformer &0.0864 & 48.3M \\
    GPS~\citep{rampasek2022GPS} & Hybrid &0.0858  &19.4M \\
    GEM-2~\citep{liu2022gem2} & Transformer &0.0793 &32.1M \\
    Global-ViSNet~\citep{wang2022visnet}& Transformer &0.0784  &78.5M \\
    Transformer-M~\citep{luo2022transformerM} & Transformer  &0.0772  &69M \\
    \midrule
    GPS++ & Hybrid &0.0778  &44.3M \\
    GPS++* & Hybrid &0.0755  &44.3M \\
    \bottomrule
    \multicolumn{4}{r}{\raisebox{-2pt}{\footnotesize{* Trained on \texttt{train+half\_valid} data split.}}} 
    \end{tabular}
    \vspace{-5pt}
\end{table}

\begin{table}
\caption{Ensembled model performance on PCQM4Mv2 dataset. Models in the proxy set are trained on the \texttt{train+half\_valid} data split whereas those in the full set are trained on all available data.}
\label{tab:results_ensemble}
\centering
\fontsize{9pt}{9pt}\selectfont
\begin{tabular}{lccccc}\toprule
\textbf{}                           & \multicolumn{3}{c}{\textbf{Proxy Set}}                          & \multicolumn{1}{c}{\textbf{Main Set}}                & \textbf{}           \\ \cmidrule(l{2pt}r{2pt}){2-4} \cmidrule(l{2pt}r{2pt}){5-5}
\textbf{}                           & \textbf{}          & \multicolumn{2}{c}{\textbf{Valid MAE}} & \textbf{}       & \textbf{Ensembling} \\ \cmidrule(l{2pt}r{2pt}){3-4} 
\textbf{Case}                       & \textbf{\# Models} & \textbf{Avg.}   & \textbf{Ensembled}   & \textbf{\# Models}          & \textbf{Weight}     \\ \midrule
1: Baseline                            & 10                 & 0.0755          & 0.0725               & 35                                          & 1                   \\
2: No Atomic Number                    & 4                  & 0.0761          & 0.0734               & 16                                            & 0.5                 \\
3: FNN Dropout = 0.412                   & 8                  & 0.0759          & 0.0729               & 14                                           & 1                   \\
4: FNN Dropout = 0.412; No Atomic Number & 5                  & 0.0761          & 0.0736               & 7                                             & 0.5                 \\
5: Feature Set 2$^\dagger$                      & 4                  & 0.0755          & 0.0731               & 15                                          & 1                   \\
6: Feature Set 3$^\dagger$                       & 4                  & 0.0754          & 0.0731               & 14                                            & 1                   \\
7: Masking Weights = {[}1,2,2{]}       & 4                  & 0.0754          & 0.0730               & 15                                            & 1                   \\ \midrule
\textbf{All}                      & \textbf{39}        & \textbf{0.0756}       & \textbf{0.0722}      & \textbf{112}                   & \textbf{}           \\
\bottomrule
\multicolumn{6}{r}{\raisebox{-2pt}{\footnotesize{$^\dagger$ As defined in Table~\ref{tab:chemical_input_features}.}}}
\end{tabular}
\vspace{-5pt}
\end{table}

\paragraph{Ensembled Model Performance} 

Diversity of models in an ensemble is a well known way to boost model performance \citep{zhou2002ensembling,lakshminarayanan2017simple}. In this work we propose six adjustments to the hyperparameters to form seven different model configurations to ensemble. In choosing these configurations we aim to build diversity in three main areas: input feature description (cases 2, 4, 5 and 6), regularisation strength (cases 2 and 3) and reliance on 3D features (case 7). As described in Section~\ref{sec:exp_setup} we build a \emph{proxy} set of models to help guide our ensembling strategy as well as a \emph{main} set trained on all available data. The details of our seven configurations and their evaluation are shown in Table~\ref{tab:results_ensemble}.

Analysing the proxy set we found that reducing the weighting for the two worst performing models gave a small improvement to our final ensembled MAE, we therefore apply this is our main final ensemble too.

The final 112 model ensemble achieves an MAE of \textbf{0.0719} on the test challenge set as reported by the competition organisers. This was completed in 1 hour 32 minutes using 1 IPU and AMD EPYC 7742 64-Core CPU including all the feature processing, program compilation and model inference, which is well under the 4 hour budget allocated.


\section{Conclusions}

In this work we define \gps, a hybrid MPNN/Transformer model, optimised for the PCQM4Mv2 molecular property prediction task~\citep{hu2021ogblsc}. Our model builds on the works of \cite{rampasek2022GPS,luo2022transformerM,godwin2022noisynodes} with a particular focus on building a powerful and expressive message passing component. We believe that the strong inductive bias induced by this part of the model is a strong driver behind the efficiency of this model which achieves performance comparable to the best published transformer while using only 64\% of the parameters.

To achieve the best overall model performance we build a diverse ensemble of 112 models and train all of these models on all the available data to form our submission to the OGB-LSC 2022 challenge. Our final \gps ensemble achieved \texttt{test-challenge} MAE of \textbf{0.0719}, placing first for this dataset.

\section*{Acknowledgements}

We would like to thank all the people from our respective organisations that have supported this work, in particular Douglas Orr, Carlo Luschi, Andrew Fitzgibbon and Ellie Dobson from Graphcore; Therence Bois and Prudencio Tossou from Valence Discovery; and Mikhail Galkin from Mila.  

\bibliographystyle{unsrtnat}
\bibliography{references}

\clearpage
\appendix
\renewcommand\thefigure{\thesection.\arabic{figure}}
\renewcommand\thetable{\thesection.\arabic{table}}


\end{document}